# Percolation pathway switching in laser graphitized polyimide conducting tracks

M J Whitfield,[1] L Yip,[a)] S Speakman,[b)] and D G Hasko
*Engineering Department, University of Cambridge*

(*Electronic mail: david.g.hasko@gmail.com)

(*Electronic mail: melanie.j.whitfield218@gmail.com)

(Dated: 12 July 2023)

Laser processing has been used to create weakly conducting tracks in polyimide film. Raman spectroscopy shows that these tracks consist of nanometre sized graphitic regions contained in a carbon-rich matrix. The measured temperature dependent and electric field dependent conduction characteristics show an activated characteristic that is consistent with nearest neighbour hopping. In addition, discrete percolation pathway switching events are seen when the system is subjected to significant disturbance from equilibrium.

Disordered electron transport is of interest for a range of device applications, including thermometry [1], ovonic memory [2], sensors [3] and information processing [4]. Many semiconductors and granular metal films show disordered transport behaviour, but engineering the transport properties requires very precise control over material composition and structure. A common feature of these disordered systems is the presence of highly conducting islands within a non-conducting matrix, where electron transport occurs through hopping over small gaps between these islands. Hopping is a tunnelling process that requires these gaps to be of very small size since the tunnelling probability decreases exponentially with distance. Control over the gap distance is probably the most important factor in engineering disordered transport behaviour.

In this work, we investigate the electron transport characteristics of laser graphitized tracks in polyimide (Kapton) fabricated under conditions designed to realise very weak electron conduction. Previous work [5-14] has concentrated on realising strongly conducting tracks made using high power laser light to degrade the polyimide at high temperature. This process results in a rapid loss of material through the liberation of gases resulting from polymer chain breakup [15]. The remaining material is usually a porous layer consisting of a mixture of (highly conductive) 2D and glassy (highly insulating) carbon-rich regions, as shown by Raman spectroscopy. A low resistance track is created when the interconnections between the 2D regions dominate the structure, as is the case with these thicker porous tracks.

Here, control of the laser processing has enabled a thin non-porous carbon-rich layer to be formed, which allows greater control over the interconnections between the 2D regions. This is done by using a short wavelength laser to limit the depth of material in which the light is absorbed (see the absorbance spectra [16]). The chosen frequency of 375nm lies just above the point at which degradation transitions from primarily photolytic to thermolytic, hence preserving the carbon rings which are known to align to form the graphitic content [17-20]. By adjusting the scan speed and laser power, and hence controlling the temperature profile, it is possible to systematically create tracks from a highly conductive state through to highly insulating.

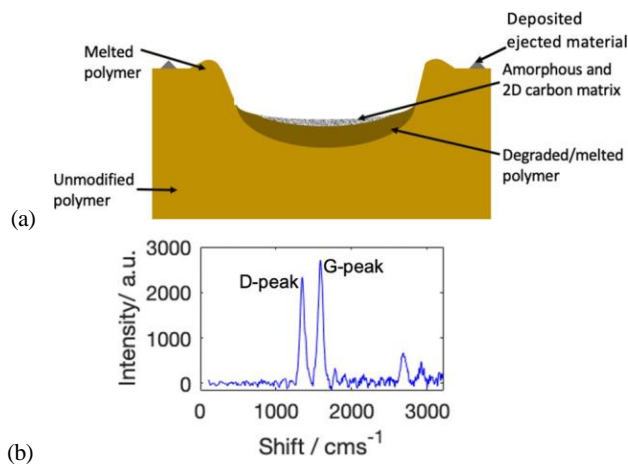

FIG. 1. (a) Vertical cross-sectional diagram of device (not to scale), (b) Raman spectra of laser carbonised conductive channel

Exposures were carried out using an L3PM tool, manufactured by 3t Technologies Limited [21]; this has a 375$nm$ laser with nominal focus diameter of 10$\mu m$. Polyimide films from a range of sources have been tested and all behave similarly. A range of laser beam powers were used with the beam turned on and off at a frequency of 30$kHz$ to realise controlled modulation (mark-to-space) ratios at various scanning speeds in order to change the exposure conditions. Higher exposure conditions (70$mW$, 50$mms^{-1}$, no modulation) realise highly conducting tracks similar to those seen in previous work. These conditions were used to form connecting leads between the channel (a single pass exposure track at 15$mW$, 75$mms^{-1}$) with length 70$\mu m$ crossing two highly conducting tracks that were contacted using indium metal away from the device under test for electrical characterisation. The laser exposed track has a carbon-rich deposit left at the base of a shallow trench a few microns in depth, as shown schematically in Fig. 1(a).

The width of this trench depends on the laser exposure conditions, but the depth remains similar across this range. The thickness of the deposit varies across the width of the track and is accompanied by significant texture differences, as seen in optical micrographs, see Table 1. Raman spectroscopy shows a ratio of peaks consistent with 2D graphitic regions of size ∼$1.4 nm$, calculated via the method given in [22], see Fig. 1(b). Assuming that the remaining material in the carbon-rich deposit is in a glassy form and so is non-conducting, then the electrical conduction mechanism will depend on the strength of coupling between the conducting 2D regions. Conduction will be confined to pathways where the gaps between these 2D regions are smallest, as shown schematically in Fig. 3(b). The conduction route with the most favourable combination of gaps for charge movement is known as a percolation pathway. Since the 2D regions will be distributed vertically as well as in the plane, the percolation pathway may be very complex. However, due to the thermal gradient with depth resulting from the laser beam exposure, the greatest density of graphitic clusters can be expected at the channel surface. The width and thickness of the trench as well as the composition of the carbon-rich deposit can be systematically controlled by changing the exposure conditions, as shown in the optical images in Table.1.

TABLE I. Optical microscopy images (x50 magnification) for tracks fabricated at $100 mms^{-1}$ at a range of power and duty ratios

| Duty Ratio | Laser Power | | | |
| --- | --- | --- | --- | --- |
| | 28mW | 42mW | 56mW | 70mW |
| 0.8 | 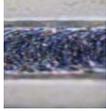 | 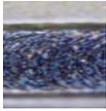 | 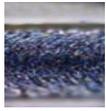 | 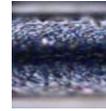 |
| 0.7 | 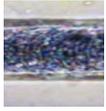 | 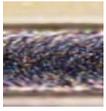 | 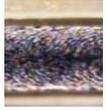 | 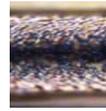 |
| 0.6 | 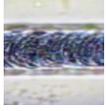 | 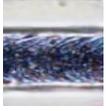 | 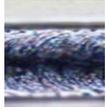 | 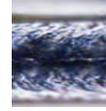 |
| 0.5 | 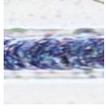 | 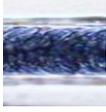 | 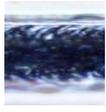 | 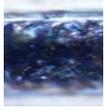[a] |

[a] Approximate field size of $10\mu m$ x $10\mu m$

Both the presence of temperature thresholds enabling thermolytic degradation, and the duration at which they are sustained, influence the resulting chemical composition and macromolecular structure. This may be observed as considerable colour and texture differences across 1) the width of the laser exposed region and 2) between optical micrographs demonstrating tracks created with different exposure scan rate, power and duty ratio, even at the same fluence. Across the conditions tested, the graphitic cluster size remains the same, but inter-cluster spacing and channel dimensions change; in this way, providing a method through which the electrical characteristics may be engineered - and weakly conducting conditions achieved.

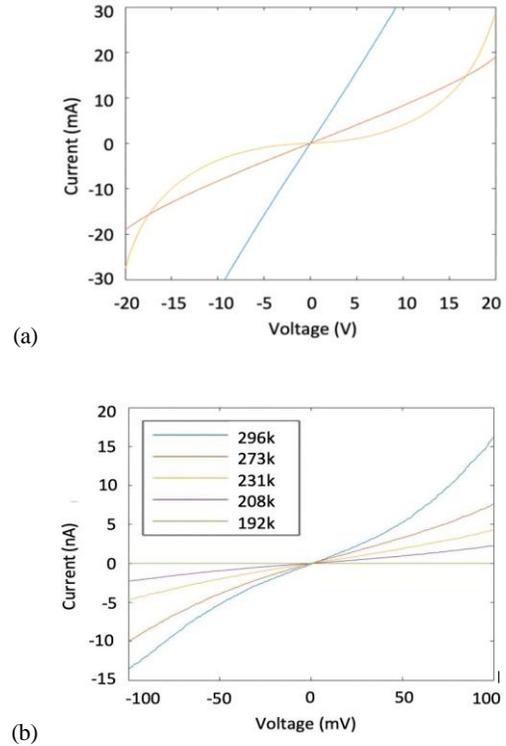

(a)

(b)

FIG. 2. (a) Three devices produced identically using the same channel irradiation conditions of $21 mW$ at $75 mms^{-1}$ and measured at 293K, (b) Device IV characteristics over temperature range (channel irradiation conditions of $15 mW$ at $75 mms^{-1}$)

As the extent of the laser exposure is reduced (by a combination of reducing the continuous laser power and by modulation) the electrical behaviour of the device becomes more non-linear, see Table 1. Under these conditions alternative conduction pathways are cut off until electrical conduction is limited to a single dominant percolation pathway within the thickest part of the deposit, where the spacing between the 2D regions is expected to be smallest. Such a weak conductor shows a significant IV non-linearity and temperature dependence, see Fig. 2(a) and (b) respectively.

In order to understand this temperature dependence, we must consider the different electron transport mechanisms in disordered systems. Dasgupta et al [23] have suggested that the electrical conductivity in amorphous carbon layers with different sp2 and sp3 content may be due to either variable-range hopping of electrons at the Fermi level near room temperature, or due to hopping in band tails, at higher temperatures. They also point out that the sp2:sp3 ratio determines the dominant transport mechanism near room temperature. A high ratio allows conduction directly by percolation just between sp2 regions; but a lower value ratio would lead to conduction being controlled by hopping through sp3 and other non-conducting materials between nearby sp2 islands. In order to distinguish between the different transport mechanisms in amorphous carbon films, Godet et al [24,25] have studied the temperature and electric field characteristics of amorphous carbon layers. They conclude that the temperature dependence of the low field conductivity and the field dependence together may distinguish most mechanisms.

The low electric field conductivity for the data shown in Fig.3(a) is given in Fig. 2(b), where the data is consistent with a simple exponential relationship with temperature i.e. an activated behaviour, which is consistent with nearest neighbour hopping. The characteristic temperature is $30K$ (for the fitted continuous curve) which is much smaller than would be expected if this temperature was due to single nm sized sp2 regions, but would be consistent with a cluster of sp2 regions together acting as an island. Such clusters would be an expected stage in the evolution towards the continuous conducting pathway as found in the low resistance wires. Hopping conduction in an organic semiconductor system has been studied by Nardes et al [26], who used a film formed by quasimetallic poly3,4-ethylenedioxythiophene (PEDOT)-rich grains embedded in continuous insulating polystyrenesulfonate (PSS) lamellas. The spin process used to create the film resulted in a wider PEDOT-rich grain separation perpendicular to the film compared to the in-plane separation. The measured temperature dependent conductivity for the in-plane direction was consistent with 3D variable range hopping, whereas the conductivity in the perpendicular direction was consistent with nearest neighbour hopping. This suggests that the spacing between the sp2 regions in our laser graphitized tracks is larger than in the other sp2:sp3 systems studied by Dasgupta et al and by Godet et al.

In order to further understand the transport mechanism we must consider the electric field dependence of the conductivity. Nardes et al [24] also derive the expected electric field dependence for their nearest neighbour hopping model, obtaining a modified conductivity (Equation 1) that includes an additional term proportional to the square of the product between the electric field and the localisation length for the perpendicular direction.

$$\sigma(F,T) \propto \sigma_0(0,T)[1 + \frac{1}{6}(\frac{eFL}{k_BT})^2] \quad (1)$$

Here the conductivity, $\sigma (F, T)$, may be related to L, the average hopping length, through change with F, the applied electric field, and T, the temperature.

The data shown in Figs. 2 (a) and (b) are the steady state readings. A typical feature observable in the initial (<5) sweeps is shown in Fig. 3 (a). In this example, the initial conductivity is high at the start voltage of -20V, but follows an erratic pathway until about -10V, where the current jumps to a well defined s-shaped characteristic. This characteristic was maintained until +10V was reached, when the current again became erratic. As the voltage reached +20V and the sweep went back towards -20V, the current traced out a new s-shaped characteristic with much lower conductivity. In subsequent sweeps the lower conductivity characteristic was maintained and this later behaviour was used to characterise the temperature dependence.

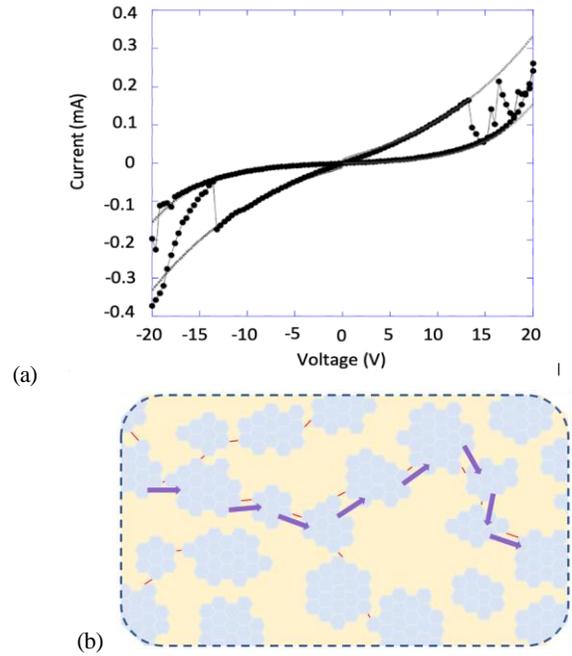

(a)

(b)

FIG. 3. (a) Current-voltage plot demonstrating pathway switching and history dependent behaviour. (b) Representation of 2D carbon clusters (blue) in a insulating matrix (yellow), possible percolation pathways are demonstrated through the red lines and the arrows indicate the single pathway responsible for conduction

Most models for the electric field dependence of hopping and related conduction mechanisms do not allow for the discontinuous IV behaviour shown in Fig. 3 (a) This behaviour shows almost discrete switching events similar in character to the random telegraph events observed in low electron number semiconductor devices. In such semiconductor devices, a switching event is seen when a trap randomly fills or empties; with a fixed voltage the current through the device switches discretely between two levels provided that the time interval between events is long compared to the current measurement time. If the switching is faster than the measurement time then the current value takes a weighted average bounded by the same two levels.

In the data shown in Fig. 3 (a), both of the non-erratic parts of the characteristic are consistent with the equations from the work of Nardes et al [24] given earlier. By fitting these two s-shaped regions separately we can see that the ohmic part of the conductivity ($R_0$) changes by a factor of almost 60 times and that the electric field constant ($E_0$) changes by a factor of ∼4. In the erratic part of the sweep, the measured current is largely bounded by these the two s-shaped characteristics, as would be expected if the switching time was much shorter than the measurement time.

The origin of this behaviour seems to be linked with the disturbance of the system at the start of the sweep. The abrupt application of the min20V at the start of the sweep seems to cause a significant disturbance from the equilibrium state, which is largely relaxed during the first sweep. This distur-bance is likely to change trap occupancies, not only along the percolation pathway, but also in nearby traps. As a result, it is possible that some of the resistances along the pathway are changed or that the route of the pathway is switched. Since at least some of this behaviour appears abrupt, a switching of the pathway seems to be the more likely explanation. One possible explanation for this switching behaviour is given by Patmiou et al [27], who consider the effects of a Poole-Frenkel mechanism combined with percolating conduction in a numerical study. They conclude that such systems will exhibit multi-valued conductivities resulting from structural modification of the percolation pathway due to the local electric field, which they call 'percolation with plasticity'. Such structural changes are the basis for operation of some types of non-volatile memory, but require significant time and energy to be dissipated in order to break and make chemical bonds, which is much less likely in carbon based materials compared to the materials considered by Patmiou et al. However, a similar change in percolation pathway may occur in low electron number systems just by the re-arrangement of trapped charge. If free electrons were confined to the sp2 regions in these laser graphitized tracks, the small size of each region (about $1nm$) would limit excess charge to a single electron per region due to Coulomb blockade (even at room temperature). Even as a cluster of sp2 regions acting as a single island, the number of free electrons would be significantly limited compared to a similar sized metal island, justifying a low electron number assumption.

In conclusion, this simple method of fabrication and the ability to engineer the conduction properties in these laser written structures allows easier implementation of devices and circuits designed to exploit this conduction mechanism. The existence of percolation pathway switching with a significant accompanying change in device resistance will enable their exploitation in novel logic and memory functions.

## A. Acknowledgements

This work was supported by the Engineering and Physical Sciences Research Council.

## B. References